\documentclass[aps,pra,twocolumn]{revtex4}
 \usepackage[usenames,dvipsnames]{color}
\usepackage{graphicx}
\usepackage{dcolumn}
\usepackage{bm}
\usepackage{amsmath}
\usepackage{amsfonts}
\usepackage{psfrag}
\usepackage[bookmarks,colorlinks=false]{hyperref}
\usepackage{mathtools}
\usepackage{accents}

\usepackage[mathcal]{euscript}

\newcommand{\braket}[2]{\langle #1|#2\rangle}
\def\bra#1{\mathinner{\langle{#1}|}} 
\def\ket#1{\mathinner{|{#1}\rangle}}

\newcommand{\Eq}[1]{Eq.~(\ref{#1})}
\newcommand{\Eqs}[2]{Eqs.~(\ref{#1})-(\ref{#2})}

\begin{document}

\title{Real-space Hopfield diagonalization of inhomogeneous dispersive media}

\author{Christopher R. Gubbin$^{1,2}$}
\author{Stefan A. Maier$^2$}
\author{Simone \surname{De Liberato}$^1$}

\affiliation{$^1$School of Physics and Astronomy, University of Southampton, Southampton, SO17 1BJ, United Kingdom}
\affiliation{$^2$Blackett Laboratory, Imperial College London, London SW7 2AZ United Kingdom}

\begin{abstract}
We introduce a real-space technique able to extend the standard Hopfield approach commonly used in quantum polaritonics to the case of inhomogeneous lossless materials interacting with the electromagnetic field. We derive the creation and annihilation polaritonic operators for the system normal modes as linear, space-dependent superpositions of the microscopic light and matter fields,  and we invert the Hopfield transformation expressing the microscopic fields as functions of the polaritonic operators. 
As an example, we apply our approach to the case of a planar interface between vacuum and a polar dielectric, showing how we can consistently treat both propagative and surface modes, and express their nonlinear interactions, arising from phonon anharmonicity, as polaritonic scattering terms.
We also show that our theory can be naturally extended to the case of dissipative materials. 
\end{abstract}
\maketitle

\section{Introduction}
Photons propagating in a polarizable medium can resonantly interact with an optically active resonance and a part of the photonic energy thus resides in the matter degrees of freedom, resulting in a dispersive reduction of the group velocity.
Quantum descriptions of such systems can be achieved through macroscopic approaches where one starts from the Maxwell equations treating the matter by a dispersive dielectric function $\epsilon\left(\omega\right)$. A modified orthonormality between the fields is derived to achieve the expected commutation relations with the final results dependant on the group velocity in the dielectric  \cite{Santos95, Milonni95}. Such approaches can describe dissipative systems by addition into the Maxwell equations of noise currents preserving the commutation relations  \cite{Knoll92}, allowing to describe the system dynamics by recourse to the Green's function of the classical scattering problem  \cite{Gruner96, Matloob95,Matloob96}. The consistency of the approach can be shown in the homogeneous  \cite{Matloob99} and general inhomogeneous case  \cite{Scheel98}. Quantization of the electromagnetic field in inhomogeneous systems is particularly relevant to the study of quantum effects in surface plasmon polaritons  \cite{Altewischer02,Fasel06,Chang06,Kolesov09,Tame13,DiMartino14} and other analogous excitations living at the interface between materials with different optical properties, like surface phonon polaritons  \cite{Borstel74} surface excitons  \cite{Lagois78}, and Tamm states  \cite{Kavokin05}. In the study of surface plasmon polaritons the quantum properties of the matter degrees of freedom are usually of little interest, and macroscopic approaches that do not explicitly account for them are very successful. Elson and Ritchie where the first to quantize the electromagnetic field at the surface of a metal described by a Drude model in the absence of dissipation  \cite{Elson71}.  This work was then extended to more general dielectric constants  \cite{Archambault10,Ballester10}. 

A different, microscopic approach to the interaction of light with matter in solid-state physics was initially pioneered by Hopfield, that considered the matter degrees of freedom as  bosonic fields coupled to the photons  \cite{Hopfield58}. The normal modes of the coupled light-matter system, usually referred to as polaritons, are then found as linear superpositions of the creation and annihilation operators of the original light and matter fields. In this way it is possible to microscopically derive results postulated in macroscopic approaches   \cite{Huttner91,Drummond99}. Losses can be taken into consideration by coupling the system to a broadband reservoir  \cite{Huttner92}. Furthermore the Hopfield approach can be made fully covariant  \cite{Belgiorno15} and equivalence can also be shown with the macroscopic Green's function approaches under the appropriate choice of dielectric function  \cite{Suttorp04}.

The Hopfield method though, has the advantage of treating on equal footing light and matter, thus becoming the tool of choice in the domain of quantum polaritonics, where nonlinear processes depending upon the matter component of the excitations are of paramount importance  \cite{Carusotto13,DeLiberato09,DeLiberato13}. Once the polaritonic operators have been obtained as linear superpositions of light and matter fields the Hopfield transformation can be inverted, allowing to express the fields describing the microscopic degrees of freedom as linear superpositions of polaritonic operators. Arbitrary nonlinearities, usually stemming from terms nonlinear in the matter field can then be naturally expressed as scattering terms between the polaritonic normal modes, allowing to investigate coherent and non-coherent polaritonic scattering processes.
Although plasmons were initially quantized by Bohm and Pines as microscopic bosonic degrees of freedom  \cite{Bohm53}, Hopfield approaches to the study of surface plasmon polaritons have appeared only recently \cite{Todorov14}, using ad-hoc methods that make it difficult to invert the Hopfield transformation. 
To the best of our knowledge, no general theory of Hopfield diagonalization in inhomogeneous materials has been formulated, and this becomes a pressing issue as inhomogeneous polaritonic systems characterised by extremely localised resonances  \cite{Caldwell13,Kim14}, and thus potentially by large nonlinear effects, become commonplace. The necessity of real-space approaches in polaritonics has also recently been highlighted in Ref. \onlinecite{Elistratov16}, proving that an inhomogeneous Hopfield theory is necessary to study sub-healing length details in polaritonic condensates.

In this work we introduce a real-space Hopfield approach to the study of non-magnetic polarizable materials, which allows to determine invertible Hopfield transformations for generic geometries. For sake of simplicity we will limit ourselves to the isotropic case, the extension to the anisotropic one not presenting any conceptual difficulty.
As an alternative to the Green's function and path integral approaches previously employed we are able to use a straightforward modal analysis in the non-dissipative case to derive the polaritonic wavefunctions, and we can normalize the resulting modes imposing bosonic commutation relations, without need to recur to flux normalization  \cite{Huttner91}. 

This paper is organised as follows: in Sec. II we introduce the general real-space Hopfield approach in the non-dissipative case. While for surface plasmon polaritons a non-dissipative treatment is accurate only at the qualitative level, 
this is not true for surface phonon polaritons, where quality factors in excess of $100$ for localised resonances are nowadays commonly achieved  \cite{Caldwell13,Chen14,Gubbin16}. The increasing interest in those systems  \cite{Greffet02,Hillenbrand02,Shen09,Holmstrom12,Li13,Ming13,Caldwell14,Feng15} and the need for a simple and powerful tool to study them has been one of the primary motivations for the present work. In Sec. III, in order to give a physical instantiation of the formal theory, we explicitly apply it to the analytically manageable case of the interface between vacuum and a polar dielectric, sustaining both propagative and surface excitations  \cite{Xu90}. 
In Sec. IV we extend the model to include losses, showing how we recover an equivalent formulation to the Green function approach of Ref. \onlinecite{Suttorp04} and the results of Ref. \onlinecite{Huttner92} in the homogeneous case.

\section{Real-space Hopfield theory for non-dissipative materials}

\subsection{Formulation of the Problem}

The coupled light-matter system can be described by the Power-Zienau-Wooley Lagrangian density  \cite{Todorov14} 
\begin{align}
\label{eq:MinCoupLag}
\mathcal{L}_0\left(\bf{r}\right)&=\int \mathrm{d}\mathbf{r}\,\,\biggr[\frac{\epsilon_0}{2}\mathrm{E}\left(\bf{r}\right)^2-\frac{1}{2\mu_0}\mathrm{B}\left(\bf{r}\right)^2 \\&\nonumber+\frac{\rho\left(\bf{r}\right)}{2}\dot{\mathrm{X}}\left(\bf{r}\right)^2  -\frac{\rho\left(\bf{r}\right) \omega_{\mathrm{T}}\left(\bf{r}\right)^2}{2}\mathrm{X}\left(\bf{r}\right)^2 \\ & -\kappa \left(\bf{r}\right) {\mathbf{X}}\left(\bf{r}\right)\cdot \dot{\mathbf{A}}\left(\bf{r}\right)-\mathrm{U}\left(\bf{r}\right)\nabla\left[\kappa \left(\bf{r}\right)\mathbf{X}\left(\bf{r}\right)\right]\biggr],\nonumber
\end{align}
where the electromagnetic field is described by the electric and magnetic components $\mathbf{E}\left(\mathbf{r}\right), \mathbf{B}\left(\mathbf{r}\right)$, related to the vector and scalar potentials $\mathbf{A}\left(\mathbf{r}\right)$ and $\mathrm{U}\left(\mathbf{r}\right)$ by
\begin{align}
\mathbf{E}\left(\mathbf{r}\right)&=-\nabla \mathrm{U}\left(\mathbf{r}\right) -\dot{\mathbf{A}}\left(\mathbf{r}\right),\\
\mathbf{B}\left(\mathbf{r}\right)&=\nabla \times \mathbf{A}\left(\mathbf{r}\right),
\end{align}
and the field $\bf{X}\left(\mathbf{r}\right)$ describes the degrees of freedom of the matter resonance with a space-dependent transverse frequency $\omega_{\mathrm{T}}\left(\bf{r}\right)$ and density $\rho\left(\bf{r}\right)$. The function $\kappa \left(\bf{r}\right)$  describes the spatially inhomogeneous light-matter coupling.
Matter is present only inside the regions where $\rho\left(\bf{r}\right)\neq0$ and thus, in order not to burden the notation, we will in the following assume that all the integrals involving matter degrees of freedom extend only inside those regions.
In the rest of this paper the spatial dependence of all variables will be suppressed where not necessary. Notice that the Lagrangian in \Eq{eq:MinCoupLag} not only models the coupling of light with microscopic harmonic degrees of freedom as excitons or phonons, but it has been shown to correctly describe plasmonic excitations in the limit of vanishing $\omega_{\mathrm{T}}$  \cite{Todorov14}.
The canonical momenta can now be calculated as
\begin{align}
\boldsymbol{\Pi}&=\frac{\delta {\mathcal{L}_0}}{\delta \dot{\mathbf{A}}}=\epsilon_0 \dot{\mathbf{A}}-\left[\kappa \mathbf{X}\right]^{\mathrm{T}},\\
\mathbf{P}&=\frac{\delta {\mathcal{L}_0}}{\delta \dot{\mathbf{X}}}=\rho\dot{\mathbf{X}},
\end{align}
where $\delta$ is a functional derivative and the superscript $\mathrm{T}$ refers to the transverse component of the field. Introducing the longitudinal frequency
\begin{align}
{\omega}_{\mathrm{L}}^2&=\omega_{\mathrm{T}}^2+\frac{\kappa ^2}{\epsilon_0 \rho},
\end{align}
we can obtain from the Lagrangian in \Eq{eq:MinCoupLag} the Hamiltonian\begin{align}
\label{eq:PZWH}
\mathcal{H}_0&=\int \mathrm{d}\mathbf{r}  \left[\frac{\mathrm{D}^2}{2\epsilon_0}+\frac{\mu_0 \mathrm{H}^2}{2}+\frac{\mathrm{P}^2}{2{\rho}}
+\frac{{\rho}{\omega}_{\mathrm{L}}^2\mathrm{X}^2}{2}-\frac{\kappa}{\epsilon_0} \mathbf{X\cdot D}\right],
\end{align}
where we expressed the electromagnetic variables in terms of the electric displacement  $\mathbf{D}=-\mathbf{\Pi}$ and magnetic field $\mathbf{H}=\mathbf{B}/{\mu_0}$.
This Hamiltonian can be quantized by imposition of canonical commutation relations, whose non-zero components read
\begin{align}
\label{eq:comm1}
\left[\mathbf{X}(\mathbf{r}),\mathbf{P}(\mathbf{r}')  \right]&=i\hbar\boldsymbol{\delta}(\mathbf{r-r'}),\\
\left[\mathbf{D}(\mathbf{r}),\mathbf{A}(\mathbf{r}')  \right]&=i\hbar\boldsymbol{\delta}^{\mathrm{T}}(\mathbf{r-r'}),
\end{align}
where $\boldsymbol{\delta}^{\mathrm{T}}(\mathbf{r-r'})$ is the transverse delta function and $\hbar$ is Planck's constant. From \Eq{eq:comm1} we can derive the further commutation relation between electric displacement and magnetic field 
\begin{equation}
\label{eq:comm2}
\left[\mathbf{D}(\mathbf{r}),\mathbf{H}(\mathbf{r}')  \right]= i\frac{\hbar}{\mu_0} \nabla' \times \boldsymbol{\delta}(\mathbf{r-r'}),
\end{equation} 
where the prime on $\nabla$ denotes it acting upon primed variables.
Being the Hamiltonian quadratic in the fields, in the spirit of the original Hopfield paper  \cite{Hopfield58}, we look for normal modes of the system in the form of polaritonic operators, linear superpositions of the light and matter microscopic fields, weighted by arbitrary space-dependent coefficients
\begin{equation}
\label{eq:Kexp}
\mathcal{K}=\int \mathrm{d}\mathbf{r} \left[\boldsymbol{\alpha}\cdot \mathbf{D} +\boldsymbol{\beta}\cdot \mathbf{H} +\boldsymbol{\gamma}\cdot \mathbf{P}+\boldsymbol{\eta}\cdot \mathbf{X}\right],
\end{equation}
where we will refer to $\boldsymbol{\alpha},\;\boldsymbol{\beta},\;\boldsymbol{\gamma},\;\boldsymbol{\eta}$ as real-space Hopfield coefficients and to $\mathcal{K}$ as the annihilation operator of a polariton mode. The choice of the four microscopic operators we use in the definition of the polariton operators in \Eq{eq:Kexp} is somehow arbitrary, and we could have used as well creation and annihilation operators for the light and matter fields, as in the original Hopfield paper  \cite{Hopfield58}, or conjugate variables $\boldsymbol{\Pi}$ and $\boldsymbol{\mathrm{A}}$ also for the electromagnetic field.
In order for the polariton operator to diagonalise the Hamiltonian $\mathcal{H}_0$ it must obey the Heisenberg equation
\begin{equation}
\label{eq:eigenvalues}
\left[ \mathcal{K}, \mathcal{H}_0\right] = \hbar\omega \mathcal{K}.
\end{equation}
Using \Eqs{eq:PZWH}{eq:Kexp} we obtain
\begin{equation}
\begin{split}
\left[ \mathcal{K},\mathcal{H}_0\right]&=i\hbar\int \mathrm{d}\mathbf{r}\, \nabla\times\boldsymbol{\alpha} \cdot\mathbf{H}
-c^2\nabla\times\boldsymbol{\beta}\cdot\mathbf{D} 
\\ & \quad  +\kappa \,c^2\nabla\times\boldsymbol{\beta}\cdot  \mathbf{X} -{\rho}{\omega}_{\mathrm{L}}^2\boldsymbol{\gamma}\cdot \mathbf{X}  +\frac{\kappa }{\epsilon_0}\boldsymbol{\gamma}\cdot \mathbf{D}
+\frac{1}{{\rho}}\boldsymbol{\eta}\cdot \mathbf{P}\\&=\hbar
\omega \int \mathrm{d}\mathbf{r}\,  \boldsymbol{\alpha}\cdot \mathbf{D}+\boldsymbol{\beta}\cdot \mathbf{H}  +\boldsymbol{\gamma}\cdot \mathbf{P}+\boldsymbol{\eta}\cdot \mathbf{X},
\end{split}
\end{equation}
which, equating the coefficients of the different operators, can be restated as the eigensystem 
\begin{align}
\label{eq:diffsys1}
\omega\boldsymbol{\alpha}&=-ic^2\nabla\times
\boldsymbol{\beta}+i\frac{\left[\kappa \boldsymbol{\gamma}\right]^{\mathrm{T}}}{\epsilon_0},\\
\label{eq:diffsys2}
\omega\boldsymbol{\beta}&=i\nabla\times
\boldsymbol{\alpha},\\
\label{eq:diffsys3}
\omega\boldsymbol{\gamma}&=i\frac{\boldsymbol{\eta}}{{\rho}},
\\
\label{eq:diffsys4}
\omega\boldsymbol{\eta}&=i\kappa \,c^2\nabla\times
\boldsymbol{\beta}-i{\rho}{\omega}_{\mathrm{L}}^2\boldsymbol{\gamma}.
\end{align}
We have thus transformed the operator-valued \Eq{eq:eigenvalues} into a set of equations, equivalent to Maxwell equations in matter, over the Hopfield coefficients. 
Such a system can be formally solved yielding discrete eigenmodes
\begin{align}
\label{eq:Psi}
 \ket{\boldsymbol{\Psi}_{{n}} }&=\left(\boldsymbol{\alpha}_{{n}} ,\boldsymbol{\beta}_{{n}} ,\boldsymbol{\gamma}_{{n}} ,\boldsymbol{\eta}_{{n}} \right),
 \end{align} 
and the relative eigenfrequencies $\omega_{{n}}$.
Bamba and Ogawa showed that polarizable matter is stable against superradiant phase transitions both in the homogeneous and inhomogeneous cases  \cite{Bamba14}, implying that  \Eqs{eq:diffsys1}{eq:diffsys4} do not present zero energy solutions. In particular they proved that, under suitable continuity conditions,  an inhomogeneous system is stable if the condition of stability for the corresponding point-wise homogeneous system is verified. In the homogeneous case, by evaluating \Eq{eq:diffsys1} and \Eq{eq:diffsys4} for $\omega=0$, we infer the stability condition $\omega_{\mathrm{T}}>0$. In the following we will thus consider a transverse frequency everywhere positive, although possibly arbitrarily small.
From direct inspection we can verify that if $\ket{\boldsymbol{\Psi}_{{n}} }$ in \Eq{eq:Psi} is solution of \Eqs{eq:diffsys1}{eq:diffsys4}  then
\begin{align}
\label{eq:Psibar}
\ket{\boldsymbol{\Psi}_{\bar{{n}}} }&=\left(\bar{\boldsymbol{\alpha}}_{{n}} ,\bar{\boldsymbol{\beta}}_{{n}} ,\bar{\boldsymbol{\gamma}}_{{n}} ,\bar{\boldsymbol{\eta}}_{{n}} \right),
\end{align} 
is also solution with eigenvalue $\omega_{\bar{{n}}} =-\omega_{{n}} $, that is for each positive energy solution there exist a negative energy one such that $\mathcal{K}_{{n}} ^{\dagger}=\mathcal{K}_{\bar{{n}}}$. 
This is a general feature, that remains valid also if we chose a different representation for the polaritonic operators (e.g., representing them as linear superpositions of creation and annihilation operators), even if in this case the coefficients of the negative energy solutions will in general not be the complex conjugate of the positive energy ones. This bipartition of the solutions into positive and negative energy subspaces is fundamental for our interpretation of the positive (negative) energy polariton operators as annihilation (creation) operators for the relative excitations. 

In order to determine the eigenmodes of \Eqs{eq:diffsys1}{eq:diffsys4} it is convenient to introduce the novel variable 
\begin{align}
\label{eq:novel}
\boldsymbol{\theta}=\boldsymbol{\alpha}+i\frac{\left[\kappa \boldsymbol{\gamma}\right]^{\mathrm{L}}}{\omega\epsilon_0},
\end{align} 
where $\mathrm{L}$ denotes the longitudinal component, that allows us to restate \Eqs{eq:diffsys1}{eq:diffsys4}, as the wave equation
\begin{align}
\label{eq:waveeq}
	\nabla \times \nabla \times \boldsymbol{\theta} = \frac{\omega^2 \epsilon\left(\omega\right)}{c^2} \boldsymbol{\theta},
\end{align}
where
\begin{align}
	\epsilon\left(\omega\right) = \frac{{\omega}_{\mathrm{L}}^2 - \omega^2}{\omega_{\mathrm{T}}^2 - \omega^2},
\end{align}
is the (generally space-dependent) dielectric function of a lossless dielectric composed of Lorentz oscillators with ${\omega}_{\mathrm{L}}$ and $\omega_{\mathrm{T}}$ as longitudinal and transverse resonant frequencies. 
We recognise in \Eq{eq:waveeq} the electromagnetic wave equation in an inhomogeneous, non-magnetic medium, that allows us to solve the differential problem in terms of $\boldsymbol{\theta}$ with the usual methods employed in classical electromagnetism. The coefficients of the polariton operator can then be calculated from \Eq{eq:novel} and \Eqs{eq:diffsys1}{eq:diffsys4} as
\begin{align}
\boldsymbol{\alpha} & = \boldsymbol{\theta}^{\mathrm{T}},\\
\boldsymbol{\beta} & = \frac{i }{\omega}  \nabla \times \boldsymbol{\theta},\\
\boldsymbol{\gamma} &= i\frac{\kappa\omega}{{\rho}\left(\omega_{\mathrm{T}}^2 - \omega^2\right)}  \boldsymbol{\theta},\\
\boldsymbol{\eta} &= \frac{\kappa \omega^2}{\omega_{\mathrm{T}}^2 - \omega^2} \boldsymbol{\theta}.
\label{eq:betaexpression}
\end{align}
Still we notice that \Eq{eq:waveeq} is generally not in the form of an eigenvalue problem, due to the simultaneous frequency and position dependence of the dielectric function, making it simpler to reason using the eigensystem in \Eqs{eq:diffsys1}{eq:diffsys4}.
Once the eigenmodes have been determined, they can be normalized requiring the polariton operators to obey
\begin{align}
\label{eq:BosonComm}
	\left[\mathcal{K}_{{m}}, \mathcal{K}_{{n}}^{\dagger}\right] &=\delta_{\mathrm{m,n}}\,\mathrm{sgn}(\omega_{{n}}),
\end{align}
where $\mathrm{sgn}(\omega)=\tfrac{\omega}{\lvert\omega\rvert}$ is the sign function. Such an equation reduces to the standard bosonic commutation relation if we restrict the indexes ${m}$ and ${n}$ only over the positive energy solutions.
In term of the Hopfield coefficients \Eq{eq:BosonComm} reads 
\begin{align}
\label{eq:BosonCommCoeff}
	&i\hbar\int \mathrm{d}\mathbf{r}\,\frac{1}{\mu_0}\boldsymbol{\alpha}_{m}\cdot\nabla\times\bar{\boldsymbol{\beta}}_{n}-\frac{1}{\mu_0}\nabla\times{\boldsymbol{\beta}}_{m}\cdot\bar{\boldsymbol{\alpha}}_{n}
	\nonumber+\boldsymbol{\eta}_{m}\cdot\bar{\boldsymbol{\gamma}}_{n}
	-\boldsymbol{\gamma}_{m}\cdot\bar{\boldsymbol{\eta}}_{n}\\&=\hbar\epsilon_0 \int \mathrm{d}\mathbf{r} 
\frac{\epsilon(\omega_{{m}})\omega_{{m}}^2-\epsilon(\omega_{{n}})\omega_{{n}}^2}{\omega_{{m}}-\omega_{{n}}}\boldsymbol{\theta}_{{m}}\cdot\bar{\boldsymbol{\theta}}_{{n}}=\delta_{\mathrm{m,n}}\,\mathrm{sgn}(\omega_{{n}}),
\end{align}
that in the case ${m}={n}$ may conveniently be expressed as
\begin{equation}
\label{eq:bosononorm}
\hbar\omega_{{n}}\epsilon_0 \int \mathrm{d}\mathbf{r}\,\epsilon\left(\omega_n\right) \nu(\omega_{{n}})\, \boldsymbol{\theta}_{{n}}\cdot\bar{\boldsymbol{\theta}}_{{n}}=\mathrm{sgn}(\omega_{{n}}),
\end{equation}
where the function $\nu(\omega)$ is given by
\begin{equation}
\label{eq:Vrat}
\nu(\omega)=1+ \frac{1}{\epsilon\left(\omega\right)} \frac{\partial \left[\epsilon(\omega)\omega\right]}{\partial\omega} = \frac{\mathrm{v}^\mathrm{G}\left(\omega\right)}{\mathrm{v}^{\mathrm{P} }\left(\omega\right)},
\end{equation}
which is equivalent to the ratio of the local group and phase velocities $\mathrm{v}^\mathrm{G}\left(\omega\right)$ and $\mathrm{v}^{\mathrm{P}} \left(\omega\right)$  \cite{Ruppin02}.

\subsection{Expressions of the microscopic fields}
We can now express the microscopic light and matter fields as linear combinations of the polaritonic modes
\begin{align}
\label{eq:elecfieldexp}
\mathbf{D}&=\sum_{n} \mathbf{f}_{n}^\mathrm{D}\mathcal{K}_{n}
=\sum'_{n}\left[ \mathbf{f}_{n}^\mathrm{D}\mathcal{K}_{n}+\bar{\mathbf{f}}_{{n}}^\mathrm{D}\mathcal{K}^{\dagger}_{n}\right],\\
\label{eq:magfieldexp}
\mathbf{H}&=\sum_{n}  \mathbf{f}_{n}^\mathrm{H}\mathcal{K}_{n}=\sum'_{n} \left[\mathbf{f}_{n}^\mathrm{H}\mathcal{K}_{n}+\bar{\mathbf{f}}_{{n}}^\mathrm{H}\mathcal{K}^{\dagger}_{n}\right],\\
\label{eq:momfieldexp}
\mathbf{P}&=\sum_{n}  \mathbf{f}_{n}^{\mathrm{P}} \mathcal{K}_{n}=\sum'_{n} \left[\mathbf{f}_{n}^\mathrm{P}\mathcal{K}_{n}+\bar{\mathbf{f}}_{{n}}^\mathrm{P}\mathcal{K}^{\dagger}_{n}\right],\\
\label{eq:polfieldexp}
\mathbf{X}&=\sum_{n}  \mathbf{f}_{n}^\mathrm{X}\mathcal{K}_{n}=\sum'_{n} \left[\mathbf{f}_{n}^\mathrm{X}\mathcal{K}_{n}+\bar{\mathbf{f}}_{{n}}^\mathrm{X}\mathcal{K}^{\dagger}_{n}\right],
\end{align}
where the primed sums are intended to be only on positive energy solutions. Expansion coefficients for the positive energy solutions can be calculated using the expansion of the polariton operators in \Eq{eq:Kexp} and the commutator relations in \Eqs{eq:comm1}{eq:comm2} as
\begin{align}
 \label{eq:ModeMag}
\mathbf{f}_{n}^\mathrm{D}&=\left[\mathbf{D},\mathcal{K}_{n}^{\dagger}\right]=-i\frac{\hbar}{\mu_0}\nabla\times\bar{\boldsymbol{\beta}}_{{n}},\\
\mathbf{f}_{n}^\mathrm{H}&=\left[\mathbf{H},\mathcal{K}_{n}^{\dagger}\right]=i\frac{\hbar}{\mu_0}\nabla\times\bar{\boldsymbol{\alpha}}_{n},\\
\mathbf{f}_{n}^{\mathrm{P}} &=\left[\mathbf{P},\mathcal{K}_{n}^{\dagger}\right]=- i\hbar \bar{\boldsymbol{\eta}}_{n},\\
\mathbf{f}_{n}^\mathrm{X}&=\left[\mathbf{X},\mathcal{K}_{n}^{\dagger}\right]= i\hbar \bar{\boldsymbol{\gamma}}_{{n}}.
\end{align}

\subsection{Orthonormality and Completeness}

It remains to verify that our procedure yields a complete set of solutions in the form of \Eq{eq:Kexp}, satisfying \Eq{eq:BosonComm}.
The polariton normalization in \Eq{eq:BosonCommCoeff} offers the natural inner product over the space of Hopfield coefficients
$\ket{\boldsymbol{\Psi}}=\left(\boldsymbol{\alpha},\boldsymbol{\beta},\boldsymbol{\gamma},\boldsymbol{\eta}\right)$, as orthonormality would assure the different polariton modes  respect the required commutator relations. We thus define the inner product between two wavefunctions 
$\ket{\boldsymbol{\Psi}'}$ and $\ket{\boldsymbol{\Psi}}$ as
\begin{align}
\label{eq:innerproduct}
\langle\braket{\boldsymbol{\Psi}'}{\boldsymbol{\Psi}}\rangle=\,
	i\hbar\int \mathrm{d}\mathbf{r}\biggr[&\frac{1}{\mu_0}\boldsymbol{\alpha}\cdot \nabla\times\bar{\boldsymbol{\beta}}'-\frac{1}{\mu_0}\nabla\times{\boldsymbol{\beta}}\cdot\bar{\boldsymbol{\alpha}}'\\&
		+\boldsymbol{\eta}\cdot\bar{\boldsymbol{\gamma}}'
	-\boldsymbol{\gamma}\cdot\bar{\boldsymbol{\eta}}'\biggr].\nonumber
\end{align}
From \Eq{eq:BosonCommCoeff} such an inner product is not positive definite and thus the vector space of the $\ket{\boldsymbol{\Psi}}$ does not form an Hilbert space over the scalar product in \Eq{eq:innerproduct}. Still it forms a Kre\v\i n space  \cite{Dritschel96} with signature operator 
\begin{align}
\eta=&\sum'_{n} \left[\ket{\boldsymbol{\Psi}_{{n}} }\rangle\langle\bra{\boldsymbol{\Psi}_{{n}} }-\ket{\boldsymbol{\Psi}_{\bar{{n}}} }\rangle\langle\bra{\boldsymbol{\Psi}_{\bar{{n}}} }\right],
\end{align}
allowing us to recover the structure of an Hilbert space over the inner product
\begin{align}
\label{eq:competenessformal}
\braket{\boldsymbol{\Psi}'}{\boldsymbol{\Psi}}=&\langle\bra{\boldsymbol{\Psi}'}\eta\ket{\boldsymbol{\Psi}}\rangle.
\end{align}
Recasting the eigensystem in  \Eqs{eq:diffsys1}{eq:diffsys4} in operatorial form as 
\begin{align}
\mathcal{B}_0\ket{\boldsymbol{\Psi}}&=\omega \ket{\boldsymbol{\Psi}},
\end{align}
we have to impose $\mathcal{B}_0$ to be self-adjoint with respect to an  inner product defined in \Eq{eq:competenessformal}, that is 
\begin{align}
\label{eq:self}
\braket{\boldsymbol{\Psi}'}{\mathcal{B}_0 \boldsymbol{\Psi}}
- \braket{\mathcal{B}_0 \boldsymbol{\Psi}'}{\boldsymbol{\Psi}} &=0.
\end{align} 
Using \Eqs{eq:diffsys1}{eq:diffsys4} and \Eqs{eq:innerproduct}{eq:competenessformal} we can transform \Eq{eq:self} into
\begin{align}
\label{eq:bcon1}
\int \mathrm{d}\mathbf{r} \left[\boldsymbol{\alpha}\cdot\nabla\times\nabla\times\bar{\boldsymbol{\alpha}}'
-\bar{\boldsymbol{\alpha}}'\cdot\nabla\times\nabla\times{\boldsymbol{\alpha}}\right]&=0.
\end{align}
Imposing the condition $\nabla\cdot{\boldsymbol{\alpha}}=0$, that from \Eq{eq:ModeMag} is just the transversality of the electric displacement field $\nabla\cdot{\boldsymbol{\mathrm{D}}}=0$, and using Green's second identity, we can put \Eq{eq:bcon1} in the form
\begin{align}
\label{eq:boundarycondition}
\int_{\partial \mathcal{V}} \mathrm{d}S\,\left[
\boldsymbol{\alpha}\cdot \left(\mathbf{n}\cdot\nabla\right) \bar{\boldsymbol{\alpha}}'-\bar{\boldsymbol{\alpha}}' \cdot\left(\mathbf{n}\cdot\nabla\right){\boldsymbol{\alpha}}\right]=0,
\end{align}
where $\partial \mathcal{V}$ is the surface of the quantization volume $\mathcal{V}$, $\mathrm{d}S$ the infinitesimal surface element, and $\mathbf{n}$ the unit vector normal to it. It is possible to satisfy \Eq{eq:boundarycondition} imposing proper homogeneous boundary conditions for $\boldsymbol{\alpha}$ on $\partial \mathcal{V}$ and under such conditions the system in \Eqs{eq:diffsys1}{eq:diffsys4} is self-adjoint, allowing us to find a complete set of solutions in the form of the polaritonic operators defined in \Eq{eq:Kexp} respecting bosonic commutation relations in \Eq{eq:BosonComm}. 
We can at this point write the completeness relation that, for arbitrary $\ket{\boldsymbol{\Psi}}$ and $\ket{\boldsymbol{\Psi}'}$, reads
\begin{eqnarray}
\bra{\boldsymbol{\Psi}}\left[ \sum_n \ket{\boldsymbol{\Psi}_n}\bra{\boldsymbol{\Psi}_n}\right]\ket{\boldsymbol{\Psi}'}= \braket{\boldsymbol{\Psi}}{\boldsymbol{\Psi}'},
\end{eqnarray}
 leading to expressions between the Hopfield coefficients of the form
\begin{align}
\label{eq:completenessexplicit}
\sum_n'\biggr[\boldsymbol{\alpha}_{{n}}(\mathbf{r'})\cdot\nabla\times\bar{\boldsymbol{\beta}}_{{n}}(\mathbf{r})
-\bar{\boldsymbol{\alpha}}_{{n}}(\mathbf{r'})\cdot\nabla\times{\boldsymbol{\beta}}_{{n}}(\mathbf{r})\biggr]=i\frac{\mu_0}{\hbar}\delta^{\mathrm{T}}(\mathbf{r-r'}),
\end{align}
allowing us to verify the consistence of our procedure by explicitly calculating the commutators we used to quantize the theory
\begin{align}
	\left[\mathbf{D}(\mathbf{r}),\mathbf{H}(\mathbf{r'}) \right]=& \frac{\hbar^2}{\mu_0^2}\sum'_{{n}}\biggr[
	\nabla\times\bar{\boldsymbol{\beta}}_{{n}}(\mathbf{r})\cdot
	\nabla'\times{\boldsymbol{\alpha}}_{{n}}(\mathbf{r'})\\&\quad\quad\;
	-\nabla\times{\boldsymbol{\beta}}_{{n}}(\mathbf{r})\cdot
	\nabla'\times{\bar{\boldsymbol{\alpha}}}_{{n}}(\mathbf{r'})\biggr]\nonumber\\
		=&\frac{\hbar^2}{\mu_0^2}\sum'_{{n}}\nabla'\times\biggr[\nonumber
	\nabla\times\bar{\boldsymbol{\beta}}_{{n}}(\mathbf{r})\cdot
	{\boldsymbol{\alpha}}_{{n}}(\mathbf{r'})\\&\quad\quad\quad\quad\;\,\,
	-\nabla\times{\boldsymbol{\beta}}_{{n}}(\mathbf{r})\cdot
	{\bar{\boldsymbol{\alpha}}}_{{n}}(\mathbf{r'})\biggr]\nonumber\\
	=&i\frac{\hbar}{\mu_0}\nabla'\times\delta^{\mathrm{T}}(\mathbf{r-r'}),\nonumber
\end{align}
with equivalent expressions arising for the other coefficients and leading to the other commutators between the microscopic fields.
The Hamiltonian in \Eq{eq:PZWH} can thus be put in the diagonal form
\begin{align}
\mathcal{H}_0=&\sum'_{{n}} \hbar\omega_{{n}}\, \mathcal{K}_{{n}}^{\dagger} \mathcal{K}_{{n}}.
\end{align}

\section{Application: Phonon Polaritons at a Planar Interface}
\begin{figure}
	\includegraphics[width=0.45\textwidth]{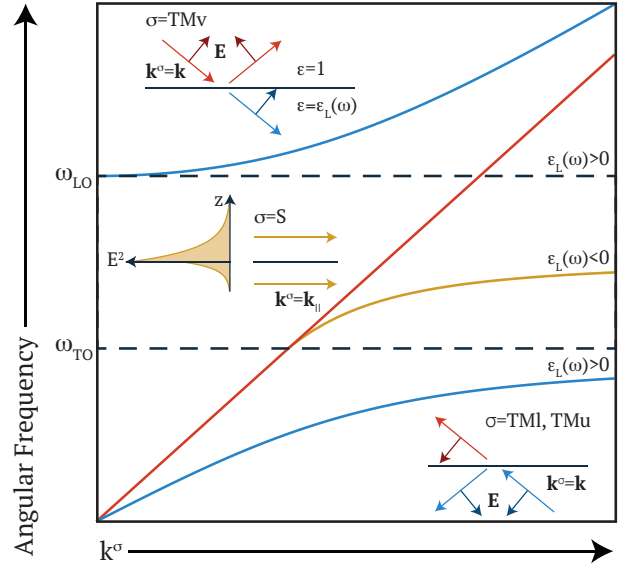}
	\caption{\label{fig:Fig1} Dispersions of the different classes of solutions of a planar interface between vacuum and a polar dielectric.
	The dispersion of photons in vacuum (red) and of bulk phonon polaritons in the dielectric (blue) are plotted as a function of the three-dimensional wavevector $k$, while the dispersion of the surface modes (yellow) is plotted as a function of the two-dimensional in-plane wavevector $k_{\parallel}$. Representative field profiles for TM bulk and surface solutions are represented as insets.}
\end{figure}
In order to illustrate the method developed in the previous section, we will apply it to the well-known case of a planar  interface between vacuum and a polar dielectric described by the Lorentz dielectric function, where the coupling of the electromagnetic field with the transverse optical phonons gives rise to both bulk and surface phonon polariton excitations \cite{Xu90}. Assuming the surface to lie in the $x-y$ plane, the air-dielectric interface, sketched in Fig.~\ref{fig:Fig1}(a), will be described by the dielectric function
\begin{align}
\label{epsvd}
\nonumber
	\epsilon\left(\omega,z>0\right) &= 1,\\
	\epsilon\left(\omega,z<0\right) &= \epsilon_{\mathrm{L}}\left(\omega\right)=  \frac{{\omega}_{\mathrm{LO}}^2- \omega^2}{\omega_{\mathrm{TO}}^2 -\omega^2},
\end{align}
where ${\omega}_{\mathrm{LO}}$ and ${\omega}_{\mathrm{TO}}$ are the longitudinal and transverse optical phonon frequencies, linked between them by the Lyddane-Sachs-Teller relation  \cite{Lyddane41}.
The eigenmodes and the relative eigenfrequencies can be calculated solving \Eq{eq:waveeq} with the dielectric function in \Eq{epsvd}, leading to both bulk and surface solutions. The bulk, propagative ones, impinging upon the surface from each side, are indexed by their polarization ($\mathrm{TM}$ or $\mathrm{TE}$) and three dimensional wavevector in the medium of origin, $\mathbf{k}=(\mathbf{k}_{\parallel},k_z)$, with $k_z>0$ for waves coming from vacuum ($\mathrm{v}$) and $k_z<0$ for waves coming from the dielectric. The former ones need also an extra index over the two bulk phonon polariton branches, lower ($\mathrm{l}$) and upper ($\mathrm{u}$) polaritons, existing in the dielectric for each value of $\mathbf{k}$. The surface solutions instead form a single branch of surface phonon polaritons, indexed by the two-dimensional in-plane wavevector $\mathbf{k}_{\parallel}$, and they only exist for $ck_{\parallel}\geq \omega_{\mathrm{TO}}$. The dispersion of the different modes can be found in Fig.~\ref{fig:Fig1}.
We can thus write the general expression of a polariton operators
\begin{align}
\label{eq:Kexpspecific}
\mathcal{K}_{\mathbf{k}^{\sigma}}^{\sigma} =&\int \mathrm{d}\mathbf{r} \biggr[ \boldsymbol{\alpha}_{\mathbf{k}^{\sigma}}^{\sigma} \cdot \mathbf{D} +\frac{i}{\omega_{\mathbf{k}^{\sigma}}} \nabla \times \boldsymbol{\alpha}_{\mathbf{k^{\sigma}}}^{\sigma} \cdot \mathbf{H} \\ &\nonumber  +  i\frac{\kappa \omega_{\mathbf{k}^{\sigma}}}{\rho\left(\omega_{\mathrm{TO}}^2 - \omega_{\mathbf{k}^{\sigma}}^2\right)} \boldsymbol{\alpha}_{\mathbf{k}^{\sigma}}^{{\sigma}} \cdot \mathbf{P}  +  \frac{\kappa \omega_{\mathbf{k}^{\sigma}}^2}{\left(\omega_{\mathrm{TO}}^2 - \omega_{\mathbf{k}^{\sigma}}^2\right)} \boldsymbol{\alpha}_{\mathbf{k}^{\sigma}}^{\sigma} \cdot \mathbf{X}\biggr],
\end{align}
where $\sigma=\lbrack$TMv, TMl, TMu, TEv, TEl, TEu, S$\rbrack$ indexes the different classes of solutions, and $\mathbf{k}^{\sigma}$ is the relative two- or three-dimensional wavevector. 
In the following we will explicitly consider only the surface phonon polariton modes ($\sigma=$S) and a single kind of bulk propagative solution ($\sigma=$TEv, the $\mathrm{TE}$ polarised modes incident from the vacuum side of the interface), as the other solutions lead to very similar expressions. Introducing the orthogonal basis vectors
$\hat{\mathbf{e}}_{z}$, $\hat{\mathbf{e}}_{\parallel}$, and $\hat{\mathbf{e}}_{\perp}$, oriented along the $z$ axis and in the $x-y$ plane respectively parallel and perpendicular to $\mathbf{k}_{\parallel}$, we can write the positive energy solutions of \Eq{eq:waveeq} for the TEv modes as
\begin{align}
\label{eq:prop}
\boldsymbol{\theta}^{\mathrm{TEv}}_{\mathbf{k}}(\mathbf{r}_{\parallel},z>0) &=  \mathrm{N}^{\mathrm{TEv}}_{\mathbf{k}} e^{-i \mathbf{k}_{\parallel}\cdot\mathbf{r}_{\parallel} } \left(e^{i k_{z} z} + r^{\mathrm{TEv}}_{\mathbf{k}} e^{-i k_{z} z}\right) \hat{\mathbf{e}}_{\perp},\\
	\boldsymbol{\theta}^{\mathrm{TEv}}_{\mathbf{k}}(\mathbf{r}_{\parallel},z<0) &=  \mathrm{N}^{\mathrm{TEv}}_{\mathbf{k}}  t^{\mathrm{TEv}}_{\mathbf{k}} e^{-i \mathbf{k}_{\parallel}\cdot\mathbf{r}_{\parallel}} e^{i \sqrt{\epsilon_{\mathrm{L}}\left(\omega_{\mathbf{k}}\right) k^2 - k_{\parallel}^2} z}  \hat{\mathbf{e}}_{\perp},\nonumber
\end{align}
where the Fresnel coefficients read
\begin{align}
	r^{\mathrm{TEv}}_{\mathbf{k}} &= \frac{k_{z} - \sqrt{\epsilon_{\mathrm{L}}\left(\omega_{\mathbf{k}}\right){k}^2-{k}_{\parallel}^2}}{k_{z} + \sqrt{\epsilon_{\mathrm{L}}\left(\omega_{\mathbf{k}}\right){k}^2-{k}_{\parallel}^2}},\\
	t^{\mathrm{TEv}}_{\mathbf{k}} &= \frac{2  k_{z}}{k_{z} + \sqrt{\epsilon_{\mathrm{L}}\left(\omega_{\mathbf{k}}\right){k}^2-{k}_{\parallel}^2}}.
\end{align}
For the S modes we have instead
\begin{align}
\label{eq:surf}
	\boldsymbol{\theta}^{\mathrm{S}}_{\mathbf{k}_{\parallel}}(\mathbf{r}_{\parallel},z>0) &=\mathrm{N}^{\mathrm{S}}_{\mathbf{k}_{\parallel}} \biggr[ \frac{1}{\sqrt{\epsilon_{\mathrm{L}}\left(\omega_{\mathbf{k}_{\parallel}}\right)}} \hat{\mathbf{e}}_{\parallel}  +   \hat{\mathbf{e}}_z \biggr] \\ & \quad \times e^{- i \mathbf{k}_{\parallel}\cdot \mathbf{r}_{\parallel}} e^{-k_{\parallel}\sqrt{\frac{-1}{\epsilon_{\mathrm{L}}\left(\omega_{\mathbf{k}_{\parallel}}\right)}} z},\nonumber\\
	\boldsymbol{\theta}^{\mathrm{S}}_{\mathbf{k}_{\parallel}}(\mathbf{r}_{\parallel},z<0) &= \frac{\mathrm{N}^{\mathrm{S}}_{\mathbf{k}_{\parallel}}}{\epsilon_{\mathrm{L}}\left(\omega_{\mathbf{k}_{\parallel}}\right)} \biggr[- \sqrt{\epsilon_{\mathrm{L}}\left(\omega_{\mathbf{k}_{\parallel}}\right)} \hat{\mathbf{e}}_{\parallel}  + \hat{\mathbf{e}}_z \biggr]\nonumber \\ & \quad \times e^{- i \mathbf{k}_{\parallel}\cdot \mathbf{r}_{\parallel}} e^{k_{\parallel}\sqrt{-\epsilon_{\mathrm{L}}\left(\omega_{\mathbf{k}_{\parallel}}\right)} z}.\nonumber
\end{align}
In \Eq{eq:prop} and \Eq{eq:surf} the $\mathrm{N}^{\sigma}_{\mathbf{k}^{\sigma}}$ coefficients are the normalization of the different modes, that can be fixed by plugging the relevant coefficients into \Eq{eq:Kexpspecific}, and then using \Eq{eq:bosononorm}.  This yields 
\begin{align}
	\mathrm{N}^{\mathrm{TEv}}_{\mathbf{k}} =& \sqrt{\frac{1}{2 \epsilon_0 \hbar\omega_{\mathbf{k}} \mathcal{V}}},\\
	\mathrm{N}^{\mathrm{S}}_{\mathbf{k}_{\parallel}} =& \sqrt{\frac{k_{\parallel}}{\epsilon_0\hbar\omega_{\mathbf{k}_{\parallel}}  \mathcal{A}} }\left[1 - \frac{\nu_{\mathrm{L}}\left(\omega_{\mathbf{k}_{\parallel}}\right)}{\epsilon_{\mathrm{L}}\left(\omega_{\mathbf{k}_{\parallel}}\right)}  \right]^{-\frac{1}{2}}\nonumber\\&\times
	\left[\frac{1}{\sqrt{-\epsilon_{\mathrm{L}}\left(\omega_{\mathbf{k}_{\parallel}}\right)}} + \sqrt{-\epsilon_{\mathrm{L}}\left(\omega_{\mathbf{k}_{\parallel}}\right)}\right]^{-\frac{1}{2}} ,
\end{align} 
where $\mathcal{A}$ is the area of the surface and $\nu_{\mathrm{L}}\left(\omega\right)$ the function obtained using $\epsilon_{\mathrm{L}}\left(\omega\right)$ into \Eq{eq:Vrat}. 

We can finally recover the expressions of the microscopic fields by inverting the Hopfield transformation using \Eqs{eq:elecfieldexp}{eq:polfieldexp}, allowing us not only to calculate physical observables, but also to naturally express nonlinear interactions in terms of scattering between polariton operators. In the specific example we are considering here nonlinearities can arise due to phonon anharmonicity, leading to $N^\mathrm{{th}}$ order nonlinear interaction Hamiltonians that can be written in the general form 
\begin{align}
\label{eq:HNL}
\mathcal{H}_{\mathrm{NL}}&=\int \mathrm{d}\mathbf{r}\, \sum_{j_1\cdots j_N=1}^3\Phi^{j_1\cdots j_{N}}\prod_{{l}=1}^{N} \mathrm{X}_{j_l},
\end{align}
where the $j$s index space coordinates and values of the nonlinear tensor $\Phi$ for some reference structure can be found in the literature at least up to the fourth order  \cite{Vanderbilt85}.
Inverting \Eq{eq:Kexpspecific} and using \Eq{eq:betaexpression} we can express the matter field as
\begin{align}
\label{eq:Xspecific}
\boldsymbol{\mathrm{X}}&=\sum_{{\sigma}}\int \mathrm{d}\mathbf{k}^{\sigma} \frac{\kappa\hbar\omega_{\mathbf{k}^{\sigma}} }{\rho(\omega_{\mathrm{TO}}^2-\omega_{\mathbf{k}^{\sigma}}^2 )}\bar{\boldsymbol{\theta}}^{\sigma}_{\mathbf{k^{\sigma}}}\mathcal{K}^{\sigma}_{\mathbf{k}^{\sigma}},
\end{align}
where $\sigma$ runs over all the different classes of solutions as in \Eq{eq:Kexpspecific}, and including both positive and negative energy solutions.
Substituting \Eq{eq:Xspecific} into the nonlinear Hamiltonian in \Eq{eq:HNL} we thus obtain the interaction Hamiltonian written as scattering terms between the polaritonic operators
\begin{align}
\mathcal{H}_{\mathrm{NL}}&=\sum_{{\sigma}_1\cdots {\sigma}_N} 
\prod_{l=1}^N \left[\int \mathrm{d}\mathbf{k}^{{\sigma}_l}\right] \Xi^{{\sigma}_1\cdots {\sigma}_N}_{\mathbf{k}^{\sigma_1}\cdots \mathbf{k}^{\sigma_N}} \prod_{l=1}^N \mathcal{K}^{{\sigma}_l}_{\mathbf{k}^{{\sigma}_l}},
\end{align}
where the scattering coefficients $\Xi$ are obtained performing the relevant integrals over the space variables.

\section{Extension to the dissipative case}
The model described in Sec. II can be extended to include dissipative effects on the same line of the original Huttner and Barnett paper  \cite{Huttner92}. This is achieved by coupling the matter field to a continuum bath of harmonic oscillators modelling the continuum in which the matter energy can be dissipated. 
Given the continuous character of the resulting spectrum, this will come at the cost of renouncing to the simple modal solution of \Eq{eq:waveeq} and we will have to solve instead a non-homogeneous wave equation. Still it is important to show that in this case our method remains viable, effectively recovering a result equivalent to Ref. \onlinecite{Suttorp04}, and reducing to Ref. \onlinecite{Huttner92} in the homogeneous case.

We thus consider the total Lagrangian $\mathcal{L} = \mathcal{L}_0 + \mathcal{L}_\text{B}$ where $\mathcal{L}_0$, from \Eq{eq:MinCoupLag}, describes the non-dissipative system and
\begin{align}
	\mathcal{L}_{\text{B}} = \int_0^{\infty} \mathrm{d} \zeta \left[ \frac{\rho\dot{\mathrm{Y}}_{\zeta}^2}{2}  - \frac{\rho \zeta^2\mathrm{Y}_{\zeta}^2}{2}  - \upsilon_{\zeta} \mathbf{X} \cdot \dot{\mathbf{Y}}_{\zeta}\right],
\end{align}
the bath of harmonic oscillators $\mathrm{Y}_{\zeta}$, indexed by their frequency $\zeta$, and coupled to the matter mode by the coupling $\upsilon_{\zeta}$. Analogously to what done in Sec. II we will assume that all the integrals over the bath degrees of freedom extend only in the regions where $\rho\neq0$.
The canonical momenta for the bath operators are found as
\begin{align}
	&\mathbf{Q}_{\zeta} = \frac{\delta \mathcal{L}_{\text{B}}}{\delta \dot{\mathbf{Y}}_{\zeta}} = \rho\dot{\mathbf{Y}}_{\zeta} - \upsilon_{\zeta}\mathbf{X}.
\end{align}
The corresponding total Hamiltonian will thus be given by $\mathcal{H} = \mathcal{H}_0 + \mathcal{H}_\text{B}$ with
\begin{align}
	\mathcal{H}_{\text{B}} =  \int \mathrm{d}\mathbf{r} \int_{0}^{\infty} \mathrm{d}\zeta \left[\frac{\rho \zeta^2\mathrm{Y}_{\zeta}^2}{2}  + \frac{\mathrm{Q}_{\zeta}^2}{2 \rho}  + \frac{\upsilon_{\zeta}}{\rho} \mathbf{Q}_{\zeta} \cdot \mathbf{X} + \frac{\upsilon_{\zeta}^2\mathrm{X}^2}{2 \rho}  \right],
\label{HB}
\end{align}
where the last term in \Eq{HB} can be included into $\mathcal{H}_0$ introducing the renormalized longitudinal frequency
\begin{equation}
	\tilde{\omega}_{\mathrm{L}}^2 = \omega_{\mathrm{T}}^2 + \frac{\kappa ^2}{\epsilon_0 \rho} + \int_0^{\infty} \mathrm{d} \zeta\, \frac{\upsilon_{\zeta}^2}{2 \rho^2} ,
\end{equation}
that includes the static shift from coupling to the bath. 
The full Hamiltonian $\mathcal{H}$ is quantized by imposition of commutation relations whose non-zero elements are those in \Eqs{eq:comm1}{eq:comm2}, in addition to 
\begin{equation}
	\left[\mathbf{Y}_{\zeta}\left(\mathbf{r}\right), \mathbf{Q}_{\zeta'}\left(\mathbf{r'}\right)\right] = i \hbar\boldsymbol{\delta}\left(\mathbf{r}- \mathbf{r}'\right)\delta\left(\zeta-\zeta'\right).
\end{equation} 
The polaritonic operators are now defined as
\begin{align}
\label{nonumber}
\mathcal{K}&=\int \mathrm{d}\mathbf{r} \biggr[\boldsymbol{\alpha}\cdot \mathbf{D} +\boldsymbol{\beta}\cdot \mathbf{H} +\boldsymbol{\gamma}\cdot \mathbf{P}+\boldsymbol{\eta}\cdot \mathbf{X} \\&\quad+ \int_0^{\infty} \mathrm{d} \zeta \left(\boldsymbol{\chi}_{\zeta}\cdot\mathbf{Q}_{\zeta} + \boldsymbol{\xi}_{\zeta}\cdot\mathbf{Y}_{\zeta} \right)\biggr],
\label{eq:Kexploss}
\end{align}
and solving the Heisenberg equation
\begin{equation}
	\left[\mathcal{K}, \mathcal{H}\right] = \hbar\omega \mathcal{K},
	\label{eq:CommFull}
\end{equation}
we obtain the system of equations
\begin{align}
\label{eq:Ndiffsys1}
\omega\boldsymbol{\alpha}&=-ic^2\nabla\times
\boldsymbol{\beta}+i\frac{\left[\kappa \boldsymbol{\gamma}\right]^{\mathrm{T}}}{\epsilon_0},\\
\label{eq:Ndiffsys2}
\omega\boldsymbol{\beta}&=i\nabla\times
\boldsymbol{\alpha},\\
\label{eq:Ndiffsys3}
\omega\boldsymbol{\gamma}&=i\frac{\boldsymbol{\eta}}{{\rho}},
\\
\label{eq:Ndiffsys4}
\omega\boldsymbol{\eta}&=i\kappa \,c^2\nabla\times
\boldsymbol{\beta}-i{\rho}{\omega}_{\mathrm{L}}^2\boldsymbol{\gamma}
+i\frac{\upsilon_{\zeta}}{\rho}\boldsymbol{\xi}_{\zeta},\\
\label{eq:Ndiffsys5}
\omega\boldsymbol{\chi}_{\zeta}&=i\frac{\boldsymbol{\xi}_{\zeta}}{\rho}-i\frac{\upsilon_{\zeta}}{\rho}
\boldsymbol{\gamma}\\
\label{eq:Ndiffsys6}
\omega\boldsymbol{\xi}_{\zeta}&=-i\rho\zeta^2 \boldsymbol{\chi}_{\zeta}.
\end{align}
Such a system may be solved using the method originally due to Fano  \cite{Fano56}, by using \Eq{eq:Ndiffsys6} to eliminate  $\boldsymbol{\chi}_{\zeta}$ and then writing the bath operator $\boldsymbol{\xi}_{\zeta}$ as a function of $\boldsymbol{\gamma}$ as
\begin{equation}
\label{eq:ydef}
	\boldsymbol{\xi}_{\zeta} = \mathcal{P}\left[\frac{\upsilon_{\zeta} \zeta^2 }{ \zeta^2-\omega^2}\boldsymbol{\gamma}\right] + \mathbf{y}\left(\omega\right) \delta\left(\zeta-\omega\right),
\end{equation}
where $\mathcal{P}$ refers to the principal part and $\mathbf{y}\left(\omega\right)$ is a frequency- and space-dependent function to be determined. Analogously to what done in Sec. II we can restate the system in \Eqs{eq:Ndiffsys1}{eq:Ndiffsys6} as an inhomogeneous wave equation
\begin{equation}
\label{eq:waveeqloss}
	\nabla\times \nabla\times \boldsymbol{\theta} - \frac{\omega^2 \tilde{\epsilon}\left(\omega\right)}{c^2} \boldsymbol{\theta} =i\omega\, \mathbf{j}\left(\omega\right),
\end{equation}
where 
\begin{align}
	\tilde{\epsilon}\left(\omega\right) = \frac{\tilde{\omega}_{\mathrm{L}}^2 - \omega^2- \mathcal{F}\left(\omega\right)}{\omega_{\mathrm{T}}^2 - \omega^2 - \mathcal{F}\left(\omega\right)},
	\label{eq:LossyDieFun}
\end{align}
is the complex dielectric function with
\begin{equation}
	 \mathcal{F}\left(\omega\right)=\mathcal{P}\left[\int_{0}^{\infty} \mathrm{d}\zeta \frac{\upsilon_{\zeta}^2\zeta^2}{{\rho^2}\left(\zeta^2-\omega^2\right)}\right],
\end{equation}
which is of the form derived by Wubs and Suttorp  \cite{Suttorp04}, and
the source current $\mathbf{j}\left(\omega\right)$ relates to the function $\mathbf{y}\left(\omega\right)$ as
\begin{equation}
	\mathbf{j}\left(\omega\right) = \frac{\upsilon_{\omega}}{{\kappa\rho}c^2}\left[\tilde{\epsilon}\left(\omega\right)-1\right]\mathbf{y}\left(\omega\right).
\end{equation} 
Solutions of \Eq{eq:waveeqloss} can now be determined for any value of $\omega$, leading to a continuous spectrum of solutions indexed by the frequency $\omega$ where the function $\mathbf{y}\left(\omega\right)$ takes the role of the normalization factor. In order to fix such a function we impose again the bosonic commutation relation for the polaritonic operators, in a continuous version of \Eq{eq:BosonComm} 
\begin{align}
	\left[\mathcal{K}_{\omega}, \mathcal{K}_{\omega'}^{\dagger}\right] & = 
	 \delta({\omega-\omega'})\mathrm{sgn}(\omega').
	\label{eq:bosonnorm2}
\end{align}
Following exactly the same steps as in Sec. II, extending it to wavevectors of the form  
\begin{align}
\ket{\tilde{\boldsymbol{\Psi}}}=\left(\boldsymbol{\alpha},\boldsymbol{\beta},\boldsymbol{\gamma},\boldsymbol{\eta},\boldsymbol{\chi}_{\zeta},\boldsymbol{\xi}_{\zeta}  \right),
\end{align}
we can write the total Hamiltonian in diagonal form as
\begin{align}
\mathcal{H}=&\hbar\int_{\omega>0} \mathrm{d}\omega \,\mathcal{K}_{\omega}^{\dagger} \mathcal{K}_{\omega},
\end{align}
where the integral extends over the positive energy spectrum of \Eqs{eq:Ndiffsys1}{eq:Ndiffsys6}.

\section{Conclusion}
We introduced a real-space Hopfield approach to the diagonalization of polarizable media, able to extend to inhomogeneous materials the standard machinery used in  the field of quantum polaritonics. Our approach allows us to  obtain explicit expressions for the quantum light and matter microscopic fields as a function of the polaritonic operators. Natural applications of this method are in the study of quantum nonlinear processes in inhomogeneous systems, where the microscopic nonlinearity, usually known as a nonlinear function of the matter field, can be expressed as scattering terms between polaritonic operators. We thus expect our theory will be an important tool in the developing fields studying quantum properties of plasmons and other surface excitations, where extremely small mode volumes can lead to strong nonlinear effects.

\section{Acknowledgments}
S.A.M. acknowledges support from EPSRC programme grants EP/L024926/1 and EP/M013812/1, plus ONR Global, the Royal Society, and the Lee-Lucas Chair in Physics.
S.D.L. is Royal Society Research Fellow and he acknowledges support from EPSRC grant EP/M003183/1.

\end{document}